\documentclass[12pt]{article}
\input amssym.tex

\usepackage{%
pstricks,%
pst-plot,%
epsf%
}
\usepackage{cite}
\usepackage{graphicx}
\usepackage{array,dcolumn}      
\usepackage{axodraw}
\newcommand{\hs}{\hspace*{0.5cm}}

\newcommand{\be}{\begin{equation}}
\newcommand{\ee}{\end{equation}}
\newcommand{\bea}{\begin{eqnarray}}
\newcommand{\eea}{\end{eqnarray}}
\newcommand{\bary}{\begin{array}}
\newcommand{\eary}{\end{array}}
\newcommand{\bit}{\begin{itemize}}
\newcommand{\eit}{\end{itemize}}
\newcommand{\ben}{\begin{enumerate}}
\newcommand{\een}{\end{enumerate}}
\newcommand{\crn}{\nonumber \\}

\newcommand{\al}{\alpha}
\newcommand{\la}{\lambda}
\newcommand{\bet}{\beta}

\newcommand{\va}{\varphi}

\newcommand{\fr}{\frac}

\newcommand{\bc}{\begin{center}}
\newcommand{\ec}{\end{center}}

\newcommand{\nn}{\nonumber}

\newcommand{\de}{\delta}

\begin{document}

\title{Symmetry Factors of Feynman Diagrams\\ for Scalar Fields}

\author{P. V. Dong\footnote{pvdong@iop.vast.ac.vn}, L. T. Hue\footnote{lthue@grad.iop.vast.ac.vn},
H. T. Hung\footnote{hthung@grad.iop.vast.ac.vn}, H. N.
Long\footnote{hnlong@iop.vast.ac.vn}, and N. H. Thao\footnote{nhthao@grad.iop.vast.ac.vn}\\
Institute of Physics, VAST, P.O. Box 429, Bo Ho, Hanoi 10000,
Vietnam}

\maketitle

\begin{abstract}
We calculate the symmetry factors of diagrams for real and complex
scalar fields in general form using an analysis of the Wick
expansion for Green's functions. We separate two classes of
symmetry factors: factors corresponding to connected diagrams and
factors corresponding to vacuum diagrams. The symmetry factors of
vacuum diagrams play an important role in constructing the
effective action and phase transitions in cosmology. In the
complex scalar field theory, diagrams with different topologies
can contribute the same, and the inverse symmetry factor for the
total contribution is therefore the sum of the inverse symmetry
factors, i.e., $1/S = \sum_i (1/S_i) $.

\end{abstract}

PACS number(s): 11.15.Bt, 12.39.St.\\


 \noindent

\section{Introduction}
In quantum field theory, physical processes are described by the
elements of the S-matrix, which are in turn given by Feynman
diagrams. One important task in calculating these diagrams is
determining their symmetry factors (see, e.g., \cite{cl}).
Fortunately, there are now various convenient computer programs
(for instance, FeynArts \cite{feynart} or QGRAF \cite{qraft}) for
constructing Feynman diagrams in different field theories. We note
that QGRAF does not work with vacuum diagrams, which play an
important role in effective theories. In a series of papers,
Kastening and coauthors \cite{kaste} developed an alternative
systematic approach for constructing all Feynman diagrams based on
considering a Feynman diagram as a functional of its graphical
elements. We stress that only real fields were considered in all
these papers, and complex fields were outside the scop

Our aim here is to derive a general formula describing the case of
complex scalar fields (it, of course, would also hold in the case
of real fields). This formula turns out to be easily understood
and is therefore very useful for those physicists who have not
developed good skills in computer disciplines. Below, we show that
the case of complex fields has very special features that are
absent in the case of real fields.

We return to our questions. What is the symmetry factor? How is it
constructed? We consider a $p$th-order expansion of the $n$-point
correlation function in a real scalar theory with the interaction
 $\mathcal{L}_{int}=(\lambda/4!)\phi^4$:
\be(1/p!)(1/4!)^p\langle
0|T[\phi(x_1)\phi(x_2)...\phi(x_n)\phi^4(y_1)
\phi^4(y_2)...\phi^4(y_p)]|0\rangle, \ee where the factor
$(i\lambda)^p$ and integrations over $y_1,\ y_2,...,y_p$ are
omitted because they are always presumed in the Feynman rules. Our
task is to count the number of different contractions giving the
same expression (corresponding to a Feynman diagram) \cite{ps}.
This number is equal to $N/D$, the number of all possible
contractions divided by the number of identical contractions. The
overall constant of the diagram then becomes
 $S^{-1}\equiv(1/p!)(1/4!)^p
N/D$. The number  $S$, called the symmetry factor of the diagram,
generally differs from unity. Further, the numerator  $N$ is a
product of p! interchanges of the vertices  $y_1,\ y_2,...,\ y_p$
and  $N_i$ self-contractions of the vertex  $y_i$ ($i=1, 2,...,p$)
and placements of contractions into this vertex. The value of
$N_i$ is $4!$ if there is no self-contraction, $4!/2$ if there is
one self-contraction (single bubble), and $4!/8$ if there are two
self-contractions (double bubble). Hence, $N=p!\prod_i N_i =
[p!(4!)^p]/[2^{s}8^{d}]$, where $s$ and $d$ are the respective
numbers of single-bubble and double-bubble vertices. Because a
double bubble contains two single bubbles, the total number of
single bubbles is $\beta=s+2d$.  We can rewrite
$N=[p!(4!)^p]/[2^{\beta}2^{d}]$.

In contrast, determining the denominator $D$ is not so easy.
Briefly, we evaluate it as follows. First, we consider the
interchange of vertex–vertex contractions. If there are $n$
contractions, then we have n! interchanges. Second, we consider
the interchange of the vertices  $y_1,...,y_p$ giving identical
contractions, i.e., an identical set of Feynman propagators. In
this case, there are $d!$  interchanges of $d$-type vertices times
 $g'$ interchanges of the remaining vertices. The result
is $D=g'd!\prod_{n=2,3...}(n!)^{\alpha_n}$, where $\alpha_n$ is
the number of vertex pairs with $n$ contractions. The symmetry
factor is given by \be
S=g'd!\prod_{n}(n!)^{\alpha_n}2^d2^{\beta}.\label{sf}\ee

Determining $g'$ is nontrivial  \cite{palmer} and sometimes leads
to significant problems. In the literature, only the symmetry
factors in the real scalar field theory with connected diagrams
are presented; those for the vacuum diagrams and also those for
the complex fields are absent. We note that vacuum diagrams have
applications in particle physics and cosmology, such as in the
effective action and phase transition (see, e.g., \cite{cl},
\cite{ps} and also \cite{huonglong} for a recent implication). A
formula for calculating the symmetry factors of such diagrams is
needed; our aim here is to explicitly derive such a formula for
the symmetry factor in the real scalar theory. We do this by
applying Wick's theorem and in the process show that the vacuum
diagrams are factored \cite{ryder} explicitly order by
perturbation theory order. We also study its generalizations to
complex scalar fields. We also list symmetry factors corresponding
to Feynman diagrams in both theories.

This work is organized as follows. In Sec. \ref{sec1}, we present
some notation. In Sec. \ref{sec2}, we formulate the symmetry
factor for the real scalar theory. In Sec. \ref{sec3}, we
generalize the formula to complex fields and also consider the
special features existing only in the complex theory. In Sec.
\ref{sec4}, we summarize our results and draw conclusions. The
appendix is devoted to Feynman diagrams and the corresponding
symmetry factors in both theories.

\section{\label{sec1}Notation}
We recall some ingredients of the $S$-matrix approach. The
time-evolution operator is given in terms of the action as
\cite{an}
 \bea U(t_1,t_2) & = & T \exp[iS_{int}(t_1,t_2,
\hat{\va})]\crn & = & N \left\{\exp\left(\fr 1 2 \fr{\de}{\de \va}
\Delta \fr{\de}{\de \va}\right)\exp[iS_{int}(t_1,t_2,
\varphi)]\right\}|_{...}, \label{sf1} \eea
 where symbol $|_{...}$
indicates that after differentiation, the classical fields $\va_i$
are replaced with the quantized ones $\hat{\va}_i$ and $T$ and $N$
denote the time-ordering and normal-ordering operators. The
$S$-matrix is the limit of the time-evolution operator as
$t_1\rightarrow-\infty$ and $t_2\rightarrow+\infty$. The
$c$-number function $\Delta (x,x')  $  (Feynman propagator) is
defined as \be \Delta (x,x') = T[\hat{\va}(x)\hat{\va}(x')] -
N[\hat{\va}(x)\hat{\va}(x')]. \label{sf1a} \ee
 The formula
\cite{an} \bea && T\left\{\prod_{i=1}^{n}F_i(\hat{\va})\right\} =
N \left\{\exp\left[\fr 1 2 \sum_i \fr{\de}{\de \va_i} \Delta
\fr{\de}{\de \va_i} + \sum_{i<k} \fr{\de}{\de \va_i} \Delta
\fr{\de}{\de \va_k}\right]\prod_{i=1}^{n}F_i(\va_i)\right\}|_{...}
 \label{sf1c}\eea is useful for our further presentation. We note
that the first term in the right-hand side of (\ref{sf1c}) is
present only in the real field theory.

 We recall that every Feynman diagram, as mentioned in the
introduction, has a symmetry factor. In \cite{cl,kaku}, it has the
form given by
 \be S
= g  2^\bet \prod_{n=2,3...}(n!)^{\al_n},
 \label{sf2}
\ee where $\al_n$ is the number of pairs of vertices connected by
$n$ identical self-conjugate lines, $\bet$ is the number of lines
connecting a vertex to itself, and $g$ is the number of
permutations of vertices that leave the diagram unchanged with
fixed external lines. We note that the factor
 $2^\bet$ comes from the factor
$1/2$ in the first term in the r.h.s of (\ref{sf1c}). We also note
that formula (\ref{sf2}) works only for connected diagrams but not
for vacuum diagrams. We derive the symmetry factor in the general
case as sketched in (\ref{sf}) for the real $\phi^4$ theory.

\section{\label{sec2}Symmetry factors in real scalar theory}

We consider the model with the interaction Lagrangian \be
\mathcal{L}_{int}^r =   \fr{\la}{4!} \phi^4.
 \label{sf4}
\ee It is well known that there is a direct connection between the
$S$-matrix elements and the Green's functions defined by the
expansion
$G(x_1,x_2,...,x_n)=\sum^\infty_{p=0}G^{(p)}(x_1,x_2,...,x_n)$
where the $p$th-order term has the form\bea
G^{(p)}(x_1,x_2,...,x_n)&=&\fr{i^p}{p!} \int_{-\infty}^{\infty}
d^4y_1 ...d^4y_p \langle 0| T[\phi(x_1)...\phi(x_n) \crn &&
\mathcal{L}^r_{int}(\phi(y_1))...
\mathcal{L}^r_{int}(\phi(y_p)]|0\rangle. \label{sf5}\eea This term
is called the
 $p$th-order Green's function. The  full Green's function
$G(x_1,..., x_n)$ contains every $n$--point diagram in the theory,
both connected and disconnected.

We recall that the four fields in $ L^r_{int}(\phi(y))$ are taken
at equal times. Applying (\ref{sf1c}) for Lagrangian (\ref{sf4}),
we obtain \be \phi^4(y) \sim T[\phi^4(y)] = N[\phi^4(y)] + 6
N[\phi^2(y)] \dot{\Delta} + 3 \dot{\Delta}\dot{\Delta},
\label{sf6}\ee where $\dot{\Delta}\equiv\Delta(y,y)$ denotes the
bubble diagram $\bigcirc$. We let $a, b$ and $c$ denote the three
terms in (\ref{sf6})
 \be a
\equiv N[\phi^4(y)],\hs b \equiv N[\phi^2(y)]\dot{\Delta}, \hs c
\equiv \dot{\Delta}\dot{\Delta}. \label{sf7}\ee Then we can
rewrite (\ref{sf6}) as \be \phi^4 \sim T[\phi^4] = a + 6 b + 3 c
.\label{sf8}\ee

 Green's function (\ref{sf5}) is invariant under permutations of
the interaction Lagrangians. Hence, the product of these
Lagrangians can be expanded as a sum of monomials in $a$, $b$ and
$c$ such that all terms  $a^p b^q c^t$ with given $p,\  q$ and $t$
are equivalent under integration. The overall coefficients of the
monomials in the expansion can be extracted  using the multinomial
formula \bea (x_1 + x_2 + \cdot \cdot \cdot + x_r)^p & =
&\sum_{p_1,p_2,...,p_r}\fr{p!}{p_1! p_2! \cdot \cdot \cdot p_r!}
x_1^{p_1} \cdot \cdot \cdot x_r^{p_r},\label{sf9} \\
\textrm{with}\hs
 p_1 + p_2 + p_3 + \cdot \cdot
\cdot +p_r & = & p. \nn\eea Equation (\ref{sf5}) then becomes \bea
G^{(p)}(x_1,x_2,...,x_n)&=& \fr{1}{p!}\left(\fr{i
\la}{4!}\right)^p\sum_{p_1 + p_2 + p_3 = p} \fr{p!}{p_1! p_2!
p_3!} \int_{-\infty}^{\infty} d^4y_1 ...d^4y_p \crn &&
\times\langle 0| T[\phi(x_1)...\phi(x_n)
a^{p_1}(6b)^{p_2}(3c)^{p_3} ]|0\rangle, \label{sf10}\eea where the
variables in the  integrand have the clear meaning
\bea
a^{p_1}b^{p_2}c^{p_3}=a(y_1)a(y_2)...a(y_{p_1})
b(y_{p_1+1})b(y_{p_1+2})...b(y_{p_1+p_2})
c(y_{p_1+p_2+1})c(y_{p_1+p_2+2})...c(y_{p}).\nn\eea

For the further presentation, we omit the summations and
integrations and represent the coefficients of $b$ and $c$ by \be
6 = \fr{4!}{2! 2!}, \hs 3 = \fr{4!}{2! 2! 2!} \label{sf11}\ee The
Green's function can then be rewritten in the form \bea
G^{(n)}(x_1,x_2,...,x_n)&=& (i\la)^p A B, \label{sf12}\eea where
\bea A &\equiv&
\fr{(4!)^{p_2}(4!)^{p_3}}{(4!)^{(p_1+p_2+p_3)}(2!)^{p_2}(2!)^{p_2}(2!)^{p_3}(2!)^{p_3}
(2!)^{p_3}p_1!p_2!p_3! }\label{sf13},\\
B &\equiv& \langle 0| T[\phi(x_1)...\phi(x_n)
a^{p_1}b^{p_2}c^{p_3} ]|0\rangle\label{sf16}\eea We note that the
$b$ associated with $p_2$ contains one bubble diagram while the
$c$ associated with $p_3$, contains two, a double bubble
$\bigcirc\!\bigcirc$. Hence if we let the $\bet$ be the number of
lines that connect a vertex to itself, then \be \bet = p_2 + 2
p_3. \label{sf14}\ee Moreover, these bubbles can be factored out
of the $T$-product in $B$ such that the $T$--operator does not act
on them: \be B=\langle 0| T[\phi(x_1)...\phi(x_n)
(N(\phi^4))^{p_1}(N(\phi^2))^{p_2}]|0\rangle
{\dot{\Delta}}^{p_2}{\dot{\Delta}}^{2p_3},\ee where the double
bubbles (as disconnected pieces) are vacuum subdiagrams. We also
note that $p_2$ and  $p_3$ simply coincide with the corresponding
 $s$ and  $d$ in the introduction.

The corresponding coefficient $A$ is interpreted as\be A =
\left[\fr{1}{(4!)^{p_1}(2!)^{p_2}p_1!
p_2!}\right]\left[\fr{1}{2^\bet (2!)^{p_3}p_3!
}\right].\label{sf15}\ee In this formula, $p_1!$ and $p_2!$ are
respective  numbers of permutations of $a$ and $b$ vertices,
similar to $p!$ in (\ref{sf5}). The $4!$ (powered $p_1$) and $2!$
(powered $p_2$) are symmetry factors (the number of permutations
of identical interaction-fields) respectively associated with $a$
and $b$, similar to $4!$ in (\ref{sf4}). In total, we obtain the
factor  $p_1! p_2! (4!)^{p_1}(2!)^{p_2}$, which is deduced as the
first factor in (\ref{sf15}). This factor can be simplified if we
use the $T$-product expansion for $B$. The second factor,
associated with the bubbles subdiagrams, is unchanged under
$T$-product: $p_3!$ is the number of permutations of $c$ vertices;
$2!$ (powered $p_3$) is the number of permutations of the two
single bubbles of any $c$ vertex;  $\bet$ was described above.

Next, to contract $B$ under the $T$-product, we refer to Eq.
(4.45) in \cite{ps}. The number of different contractions that
give the same expression is the product of four types of factors.
First, we have  $p_1!p_2!$ interchanges of $p_1$ $a$  and $p_2$
$b$ vertices. Second, for the placement of contractions into a
vertex, we have 4! for a and 2! for b vertices and therefore
 $(4!)^{p_1}(2!)^{p_2}$ for $p_1$ $a$ and $p_2$ $b$ vertices
 (we note that there is no
self-contraction for each vertex). Third, we have $1/\prod_{n=2,
3...}(n!)^{\alpha_n}$ interchanges of vertex-vertex contractions,
where $n$ is the number of contractions and $\alpha_n$ is the
number of vertex pairs with $n$ contractions. Finally, if we let
$g'$ be the number of interchanges of  $a$ and $b$ vertices that
do not change the diagram topologically, then the factor $1/g'$
should be multiplied to the result. In summary, the total factor
contributing to one diagram is \be \fr{p_1! p_2!
(4!)^{p_1}(2!)^{p_2}}{g' \prod_n (n!)^{\al_n}} A = \fr{1}{(g'p_3!)
2^\bet (2!)^{p_3} \prod_n (n!)^{\al_n}}\label{sf17}\ee Hence, the
symmetry factor is given by
 \be S = g 2^\bet (2!)^{d}
\prod_n (n!)^{\al_n} \label{sf18},\ee where $d=p_3$, and
$g=g'p_3!$ has the same meaning as $g'$. We note that any vertex
of $a$ and $b$ directly connected to the external points $x_1,
x_2,..., x_n$ is not subject to the interchanges defining $g'$.
The examples in \cite{palmer} and the followings examples
demonstrate this.

The constructed diagram typically consists of connected pieces
(subdiagrams), a piece connected to $x_1$, $x_2$,..., $x_n$ and
several pieces disconnected from all the external points, vacuum
bubbles, in which the double bubble is one of the cases. We let
 $V_c$ denote the connected piece and  $V_k$
denote the various possible disconnected pieces:
 \bc
\begin{picture}(200,50)(0,40)
\Text(5,50)[]{$V_k\in$}
 \GCirc(30,60){5}{1} \GCirc(30,50){5}{1}\Text(50,45)[]{,}

\GCirc(70,60){5}{1} \GCirc(70,50){5}{1} \GCirc(70,70){5}{1}
\Text(90,45)[]{,}

\GCirc(120,60){15}{1}
 \CArc(105,60)(21.2132,-45,45)
  \CArc(135,60)(21.2132,135,225)\Text(150,45)[]{,}

\Text(165,55)[]{...}
\end{picture}\ec
where $k=1,2,3...$ We suppose that the diagram has $n_k$ pieces of
the form $V_k$ for each $k$ in addition to $V_c$. Let the value of
$g$ for the connected piece $V_c$ and disconnected pieces $V_k$ be
$g_c$ and $g_k$. It is easy to obtain $g=\prod_l n_l !
(g_l)^{n_l}$, where $l=c,\ k$ and $n_c=1$, $n_1=p_3$. Here, $n_k
!$ is the symmetry factor coming from interchanging the $n_k$
copies of $V_k$. We can therefore rewrite (\ref{sf18}) as \be
S=\prod_l n_l! (S_l)^{n_l}=S_c\times\prod_k n_k!
(S_k)^{n_k},\label{sfger}\ee where $S_l=S_c,\ S_k$ is the symmetry
factor of $V_l$ having the same form as (\ref{sf18}): \be S_l=g_l
2^{\bet_l} (2!)^{d_l} \prod_{n} (n!)^{\al^l_{n}},\label{sfcon}\ee
where the parameters indexed by $l$ are those of $V_l$ satisfying
$d_{l=1}=1$, $d_{l\neq 1} =0$, $d=\sum_l n_l d_l$, $\beta=\sum_l
n_l \beta_l$ and $\al_n=\sum_l n_l \al^l_n$. We note that there is
an additional factor  $2!$ associated with only double bubble.
This  contradicts  formula (\ref{sf2}), which is given in the
literature.

In calculating, we note that the symmetry factor of arbitrary
diagrams is obtained from (\ref{sf18}) or (\ref{sfger}) while that
of connected diagrams is given by (\ref{sfcon}). Because
(\ref{sf18}) and (\ref{sfcon}) have the same form, we can commonly
use (\ref{sf18}) for both the cases with the corresponding
interpretation of the parameters. The symmetry factors of some
two-point connected diagrams are
 \bc
\begin{picture}(200,300)(40,-100)
\Line(10,170)(90,170) \GCirc(50,180){10}{1} \Text(220,175)[]{$S=2\
(g=1,\beta=1,d=0,\al_n=0)$}

\Line(10,140)(40,140) \GCirc(50,140){10}{1} \Line(40,140)(60,140)
\Line(60,140)(90,140) \Text(220,140)[]{$S=6\
(g=1,\beta=0,d=0,\al_3=1)$}

\Line(10,90)(90,90) \GCirc(35,100){10}{1}\GCirc(65,100){10}{1}
\Text(220,95)[]{$S=4\ (g=1,\beta=2,d=0,\al_n=0)$}

\Line(0,40)(100,40) \GCirc(20,50){10}{1}\GCirc(80,50){10}{1}
\GCirc(50,50){10}{1} \Text(220,45)[]{$S=8\
(g=1,\beta=3,d=0,\al_n=0)$}

\Line(10,-20)(90,-20) \GCirc(35,-10){10}{1}\GCirc(65,-10){10}{1}
\GCirc(65,10){10}{1} \Text(220,-10)[]{$S=8\
(g=1,\beta=2,d=0,\al_2=1)$}

\Line(10,-80)(90,-80)
\GCirc(35.8579,-55.8579){10}{1}\GCirc(50,-70){10}{1}
\GCirc(64.1421,-55.8579){10}{1} \Text(220,-70)[]{$S=8\
(g=2,\beta=2,d=0,\al_n=0)$}

\end{picture}\ec

For some vacuum bubbles, we also have
\bc
\begin{picture}(200,170)(40,30)
 \GCirc(30,180){10}{1}\GCirc(50,180){10}{1} \Text(220,180)[]{$S=8\
(g=1,\beta=2,d=1,\al_n=0)$}

\GCirc(20,140){10}{1}\GCirc(40,140){10}{1} \GCirc(60,140){10}{1}
 \Text(220,140)[]{$S=16\
(g=2,\beta=2,d=0,\al_2=1)$}

\GCirc(40,100){15}{1}
 \CArc(40,85)(21.2132,45,135)
  \CArc(40,115)(21.2132,225,315) \Text(220,100)[]{$S=48\
(g=2,\beta=0,d=0,\al_4=1)$} \GCirc(40,50){15}{1}
 \CArc(25,50)(21.2132,-45,45)
  \CArc(55,50)(21.2132,135,225)\GCirc(20,50){5}{1} \Text(220,50)[]{$S=24\
(g=2,\beta=1,d=0,\al_3=1)$}
\end{picture}\ec
For general diagrams, we consider the  example
 \bc
\begin{picture}(100,110)(0,-10)
 \GCirc(70,55){10}{1}
   \GCirc(40,80){10}{1} \GCirc(40,60){10}{1}
\GCirc(100,80){10}{1} \GCirc(100,60){10}{1} \Line(20,45)(120,45)
\Text(55,10)[]{$S=2.2!(8)^2=256$, using (\ref{sfger}).}
\Text(55,-5)[]{Alternatively, $S=256\ (g=2, \beta=5, d=2,
\al_n=0)$, from (\ref{sf18}).}
\end{picture}\ec

More examples of symmetry factors are given in the following
sections. In what follows, if some parameter has its trivial value
(such us $g=1$ or  $\beta=0$), then that parameter is not listed
in parentheses. We next consider the case of complex scalar
fields.

\section{\label{sec3}Symmetry factors in complex scalar theory}

The interaction Lagrangian in the complex scalar theory is
 \be \mathcal{L}_{int}^c =
\fr{\rho}{4} (\varphi^* \varphi)^2
 \label{sf22}
 \ee
Applying (\ref{sf1c}), we obtain
 \be (\varphi^* \varphi)^2 \sim
T[(\varphi^* \varphi)^2] = N[(\varphi^* \varphi)^2] + 4
N(\varphi^* \varphi) \dot{\Delta} + 2 \dot{\Delta}\dot{\Delta},
\label{sf23}\ee where $\dot{\Delta}$ in this case denotes the
bubble diagram  \emph{with arrow }
\begin{picture}(12,12)(0,0)\ArrowArc(4,4)(6,-90,270)\end{picture}.
As before, we let  $a,\ b $ and $c$ denote the corresponding
terms. The $p$th-order Green's function is\bea
G^{(p)}(x_1,x_2,...,x_n)&=& (i\rho)^p A_c
 \langle 0|
T[\va(x_1)...\va^*(x_n)  a^{p_1}b^{p_2}c^{p_3} ]|0\rangle,
\label{sf23}\eea where the integrations and summations are
understood and \bea A_c \equiv \fr{1}{4^{p_1}2^{p_3}p_1!p_2!p_3!
}.\label{sf24}\eea We note that the Green's function is nonzero
only if the number of fields
 $\va$ in (\ref{sf23}) is  equal to the number of  their
complex-conjugate fields $\va^*$.

Repeating the previous analysis, we obtain the contribution for
one diagram \be \fr{p_1 ! p_2 ! 4^{p_1}}{g' \prod_n (n!)^{\al_n}}
A_c = \fr{1}{(g'p_3!) 2^{p_3}\prod_n (n!)^{\al_n}}.\label{sf25}\ee
Hence, the symmetry factor in the theory under consideration is
given by \be S = g 2^{d} \prod_n (n!)^{\al_n}, \label{sf26}\ee
where $d=p_3$ is the number of double bubbles, and $g=g'p_3 !$ is
the number of interchanges of interacting vertices leaving both
the diagram and  its charged scalar flows unchanged. As before, we
can separate the symmetry factor  into  subfactors corresponding
to connected and vacuum subdiagrams: \be S = S_c \times S_v.
\label{sf27}\ee The symmetry factor for these subdiagrams has the
same form as (\ref{sf26}), where $d$ is nonzero only if it is
associated with a double bubble.

We emphasize that there is no factor $2^\bet$ in (\ref{sf26}). We
note that $n$ is the number of identical lines connecting two
separated vertices with {\it the same direction}. Formula
(\ref{sf26}) is simply  a generalization of (\ref{sf18})
discriminating between the  scalar field directions.  We
illustrate this with  the  examples \bc
\begin{picture}(200,100)(-20,80)
\ArrowLine(-100,140)(-70,140)\ArrowLine(-70,140)(-40,140)
\ArrowArc(-70,150)(10,-90,270) \Text(-70,100)[]{$(a)\ S=1$ }
 \ArrowLine(40,140)(60,140)
  \ArrowLine(20,140)(40,140)
   \ArrowLine(60,140)(80,140)
  \ArrowArc(50,140)(10,0,180)
    \ArrowArc(50,140)(10,180,0)
  \Text(50,100)[]{$(b) \ S=2\ (\al_2=1)$ }
\ArrowLine(155,120)(235,120) \ArrowArc(195,150)(20,0,180)
\ArrowArc(195,150)(20,180,360)
 \ArrowArc(195,130)(28.2843,45,135)
  \ArrowArc(195,170)(28.2843,225,315)
  \Text(195,100)[]{$(c)\ S=8\ (g=2,\al_2=2)$ }
 \end{picture}
 \ec
In the diagram (a) the symmetry factor is 1 because $\beta$ is
zero. In (b), we have only one set $n=2$ and  in (c),  we have
{\it two} sets with $n=2$.  We recall  that in the real scalar
theory,  we have  $n=3$ and $n=4$  for the  corresponding
diagrams. Many comparisons of symmetry factors of third-order
diagrams in the real and complex scalar theories are given in the
appendix.

It follows from Eq.(\ref{sf27}) that the vacuum diagrams are
factored order by perturbation theory order. Hence, the connected
Green's functions, as in the literature, can be defined by the
formula
 \bea < 0|
T[\varphi(x_1)\cdot \cdot \cdot \varphi(x_n)] | 0>_c = \fr{\langle
0|T[\varphi(x_1)\cdot \cdot \cdot \varphi(x_n) \exp{i \int d^4 y
\mathcal{L}_{int}}] |0\rangle}{\langle 0| T\exp{i \int d^4 y
\mathcal{L}_{int}}|0\rangle},
 \label{gr56mt}  \eea
where the vacuum diagrams are contained in the denominator.

We next discuss some special properties of the complex theory. We
consider two contributions with the symmetry factors

\bc
\begin{picture}(500,130)(0,20) \ArrowArc(93,80)(
47.124,0,60)\ArrowArc(139,80)(47.124,120,180)
\ArrowArc(116,120)(47.124,240,300) \ArrowLine(140,80)(93,80)
\ArrowLine(93,80)(116,120) \ArrowLine(116,120)(140,80)
\Text(120,50)[]{$S_1=6\ (g =3!)$ }

\ArrowArc(293,80)(47.124,0,60)\ArrowArc(339,80)(47.124,120,180)\ArrowArc(316,120)(47.124,240,300)
\ArrowLine(293,80)(340,80)\ArrowLine(316,120)(293,80)
\ArrowLine(340,80)(316,120) \Text(320,50)[]{$S_2=24\ (g
=3,\al_2=3)$}
\end{picture}\ec
It is easy to verify that these contributions coincide because
 $\Delta(x,y)=\Delta(y,x)$ \cite{greiner}.
Hence, contributions of this type can be determined by only one
diagram with the symmetry factor given by
 \be
S^{-1}=S^{-1}_1+S^{-1}_2,\ee and therefore  $S=24/5$.

We note the recently proposed hybrid inflationary scenario
\cite{linde91} in which there are two scalar fields  $\phi$ and
$\varphi$ with the coupling
 \be \fr \la 2 (\phi^2
\varphi^2).\label{sf23a} \ee It is easily to verify  that our
formula is applicable  to such interactions.

\section{\label{sec4}Conclusion}

We have derived the symmetry factor for both the real and the
complex scalar theories: \be S=g 2^\beta 2^d \prod_n
(n!)^{\al_n},\ee
 where $g$ is the  number of interchanges of vertices
leaving the diagram topologically unchanged, $\beta$ is the number
of lines connecting a vertex to  itself ($\beta$ is zero  if the
field is complex), $d$ is the number of double bubbles, and
$\al_n$ is the number of vertex pairs connected by $n$-identical
lines. Our result revises the usual symmetry factor formula in the
literature. Our result is easily generalized to higher-spin
fields.

We have also showed  that in the complex scalar theory, diagrams
with different topologies can contribute the same. We also
obtained the symmetry factor for contributions of such type.

It is easy to verify that our results are consistent with the
symmetry factors in \cite{kaste}.

 Our result explicitly shows that
the vacuum diagrams, as expected, are factored order by
perturbation theory order.

We recall that determining the symmetry factor is important
because it not only is an important component of modern quantum
field theory but also is used to calculate the effective potential
in higher-dimensional theories and cosmological models.

\section*{Acknowledgments}
 This work was supported in part
by the National Foundation for Science and Technology Development
(NAFOSTED)  under grant  No: 103.01.16.09.
\\[0.3cm]

\appendix

\section{\label{app1}Symmetry factors for three-order diagrams in the real and the complex scalar
theories}

Diagrams of the real scalar theory are given on the left below.
The corresponding diagrams of the complex scalar theory are given
on the right.

\begin{figure}[h]
\bc
\includegraphics{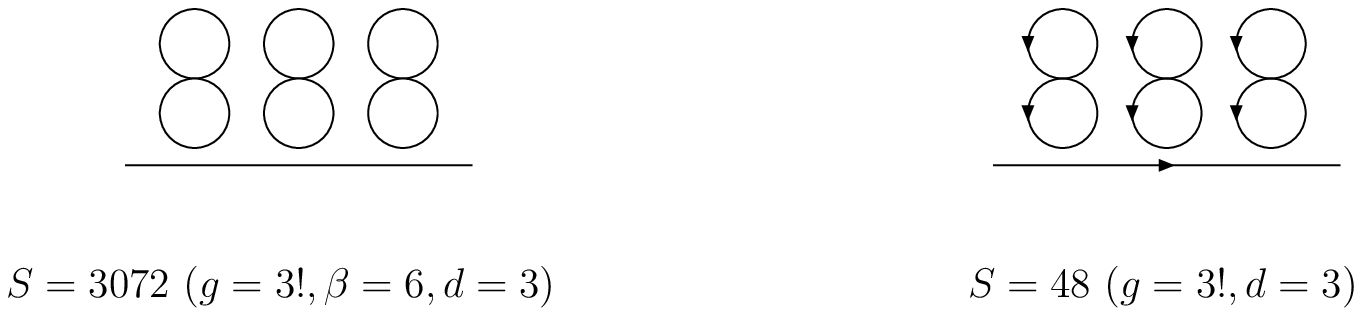}
\includegraphics{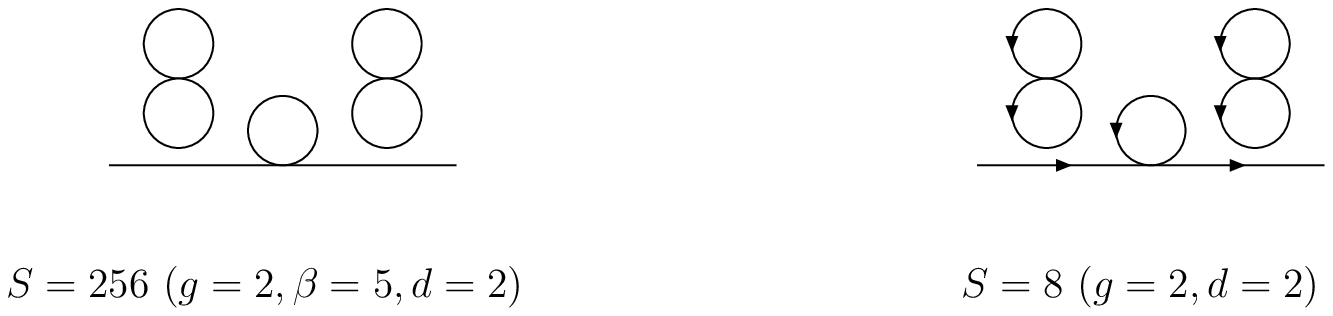}
\includegraphics{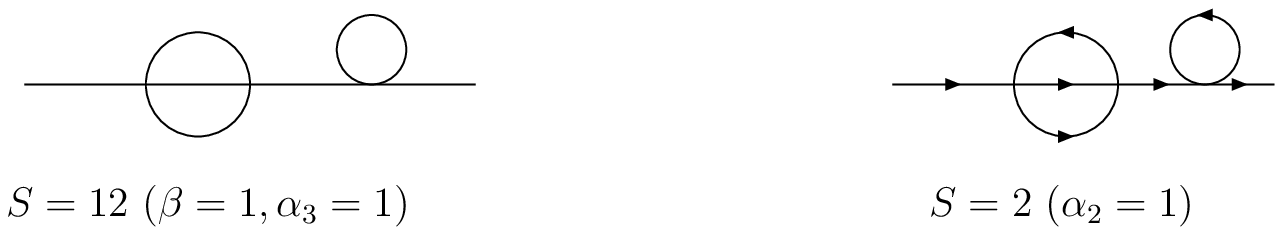}
\ec
\end{figure}

\begin{figure}[h]
\bc
\includegraphics{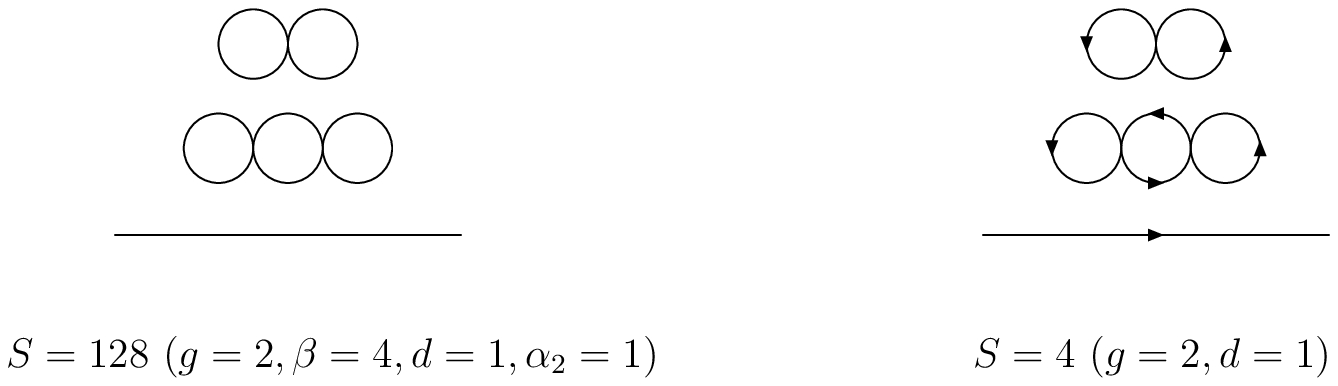}
\includegraphics{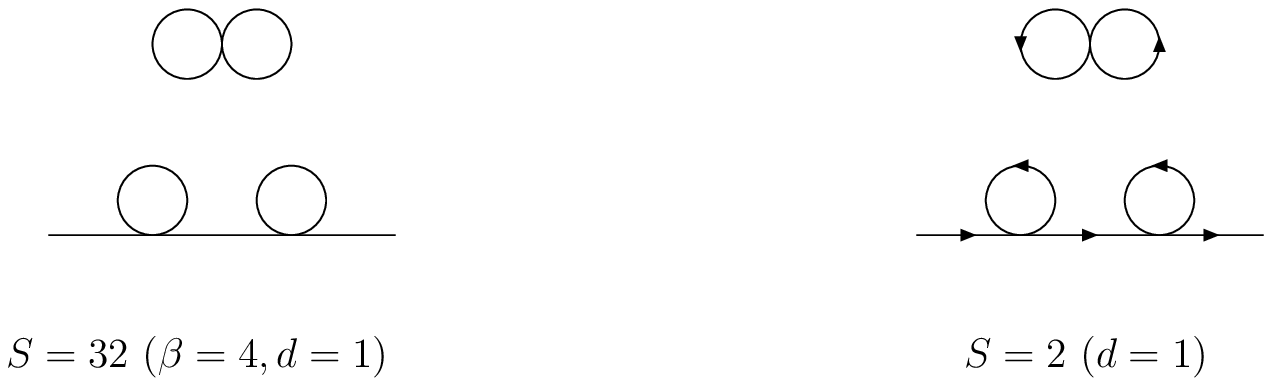}
\includegraphics{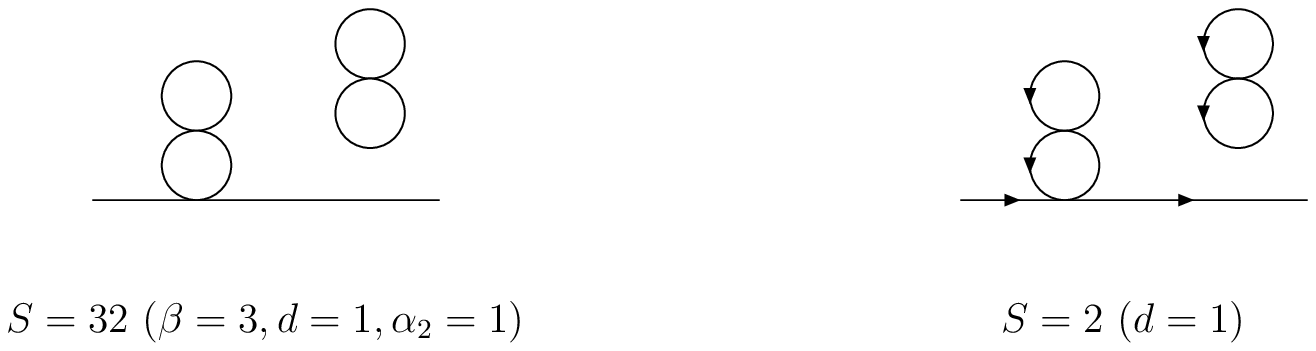}
\includegraphics{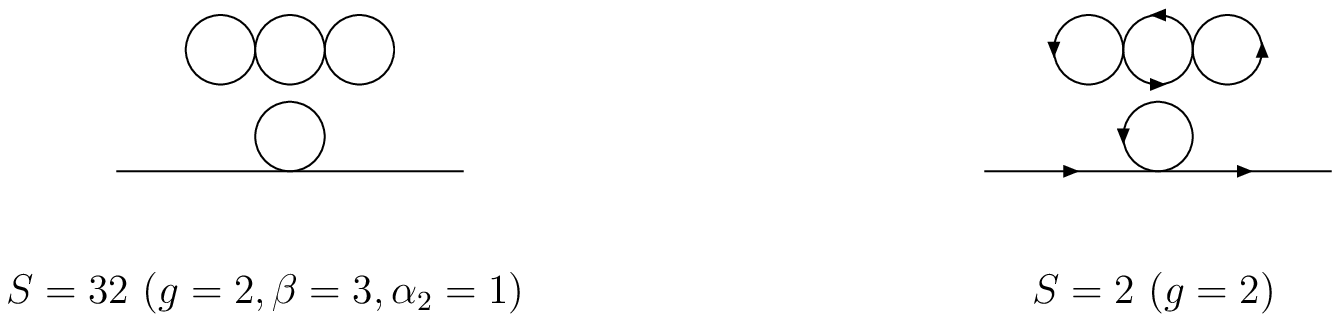}
\ec
\end{figure}

\begin{figure}[h]
\bc
\includegraphics{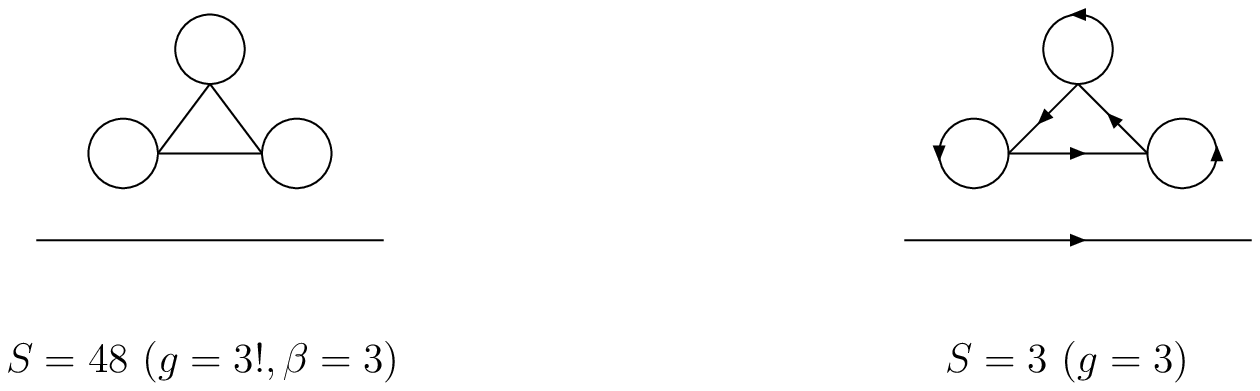}
\includegraphics{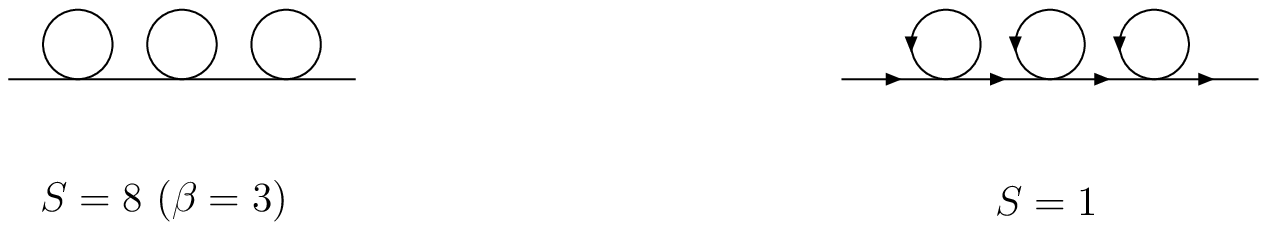}
\includegraphics{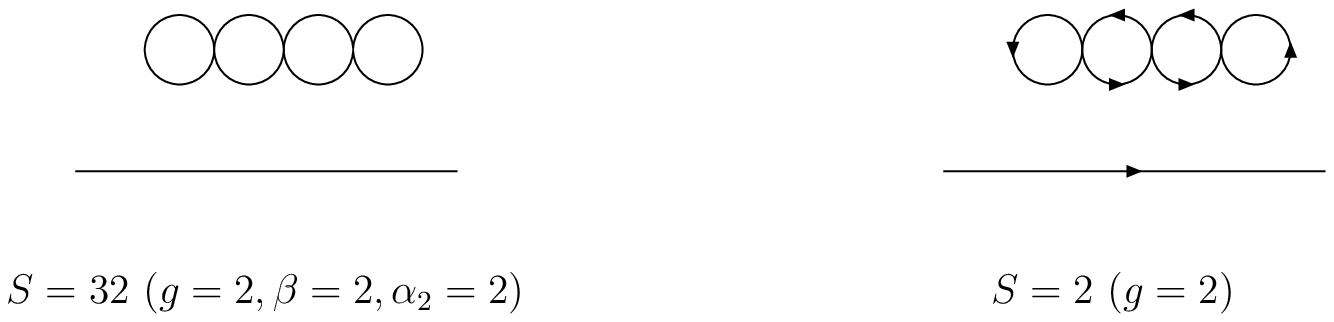}
\includegraphics{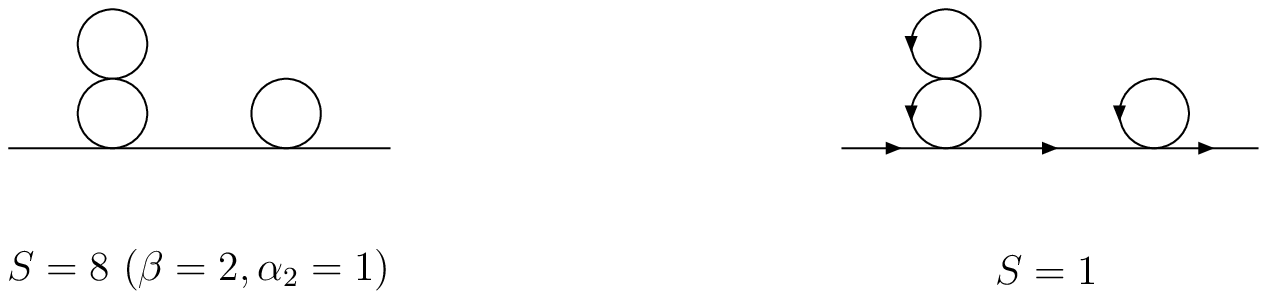}
\ec
\end{figure}

\begin{figure}[h]
\bc
\includegraphics{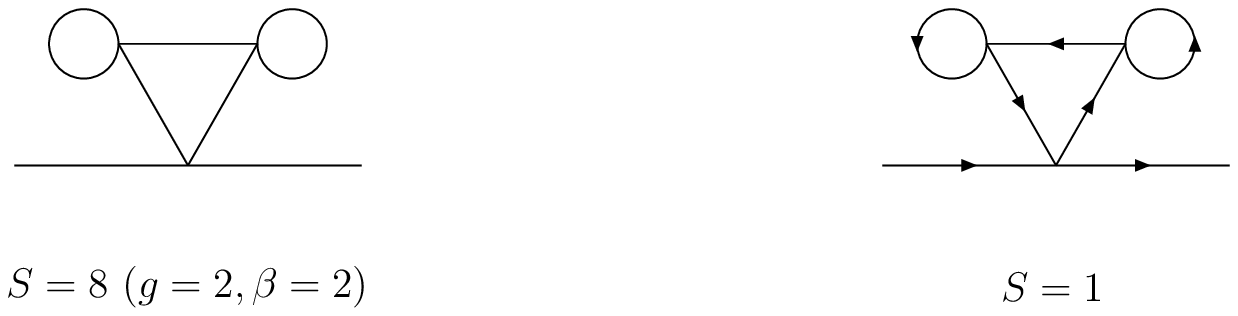}
\includegraphics{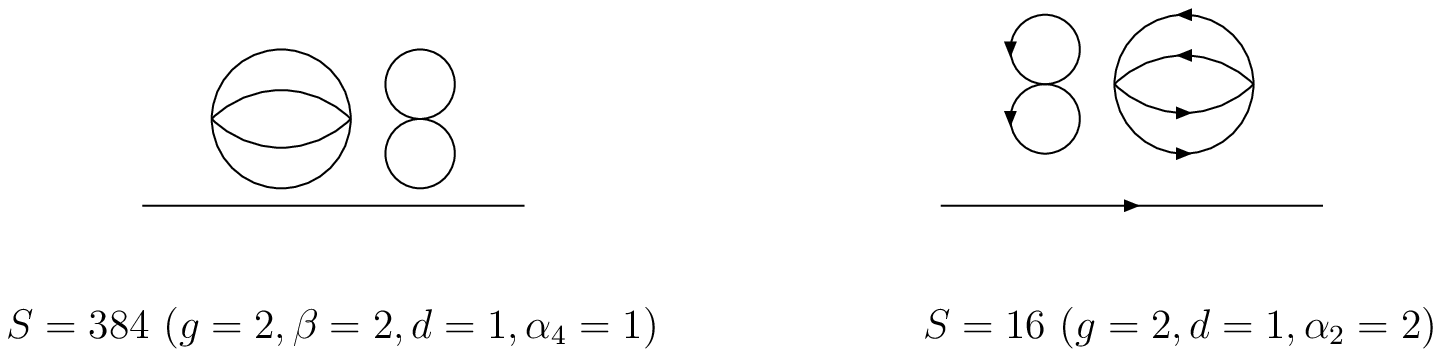}
\includegraphics{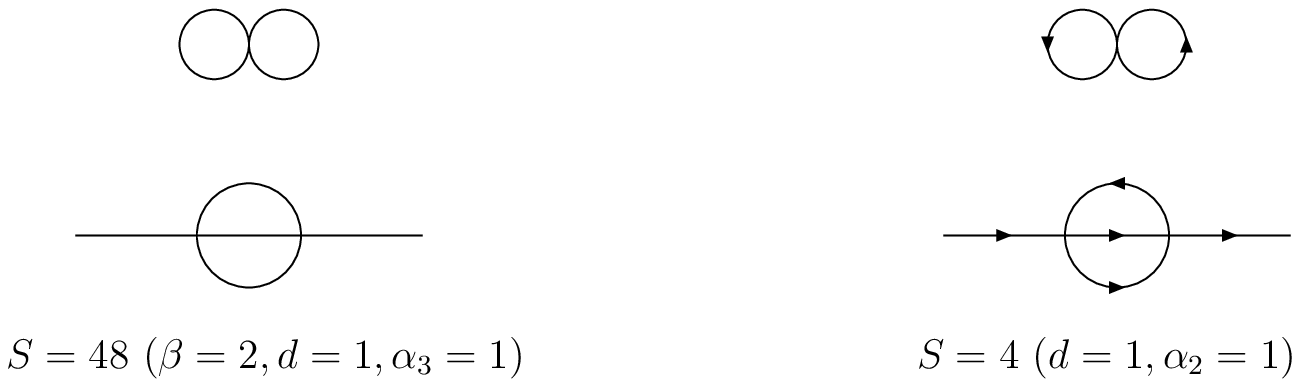}
\includegraphics{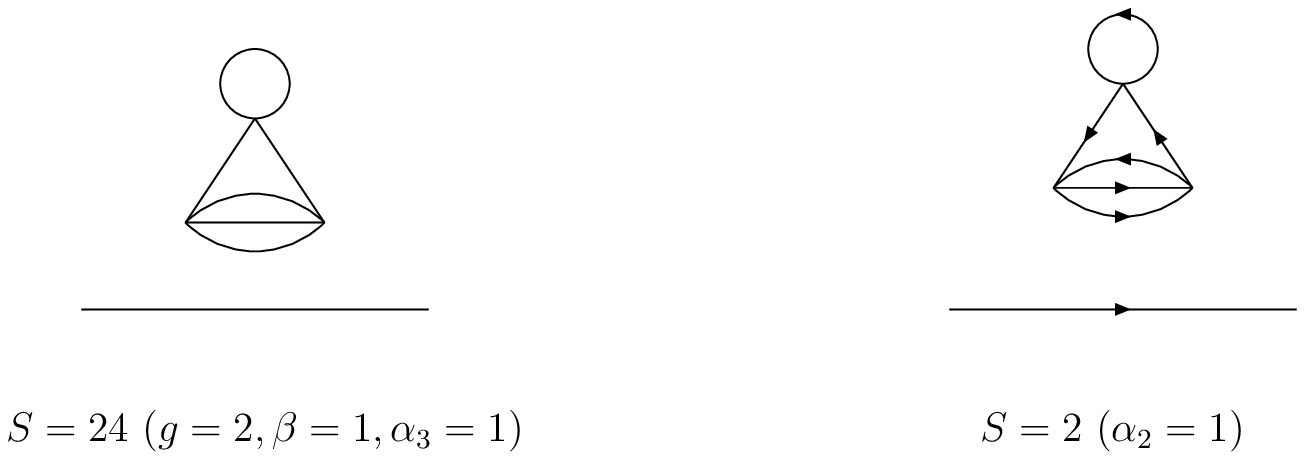}
\ec
\end{figure}

\begin{figure}[h]
\bc
\includegraphics{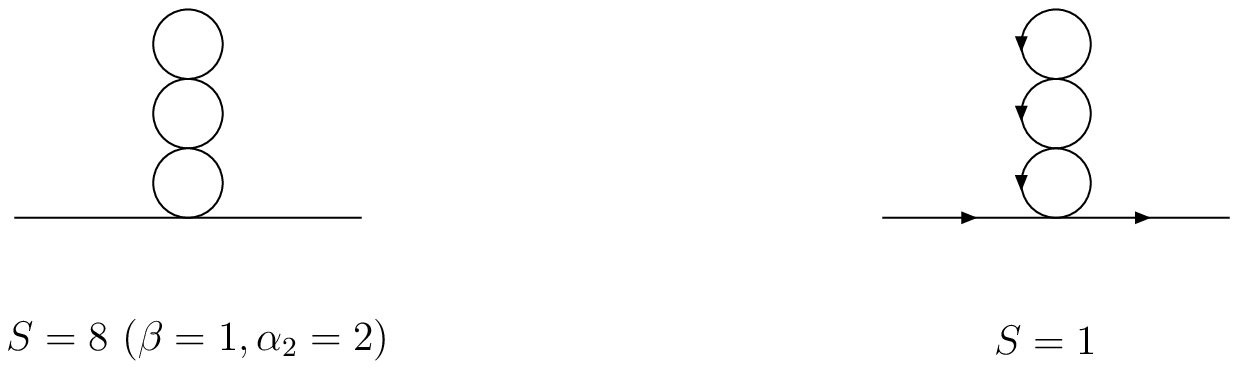}
\includegraphics{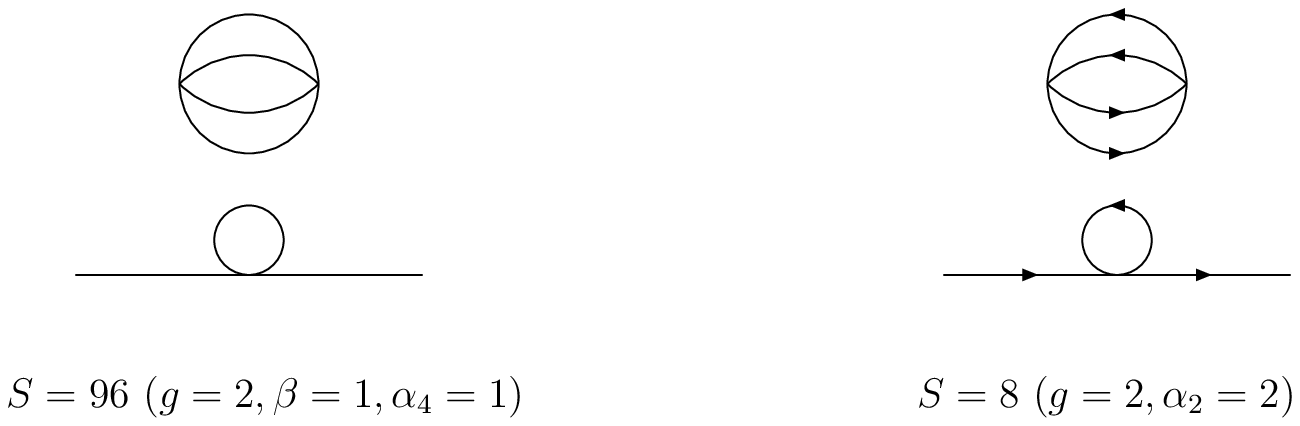}
\includegraphics{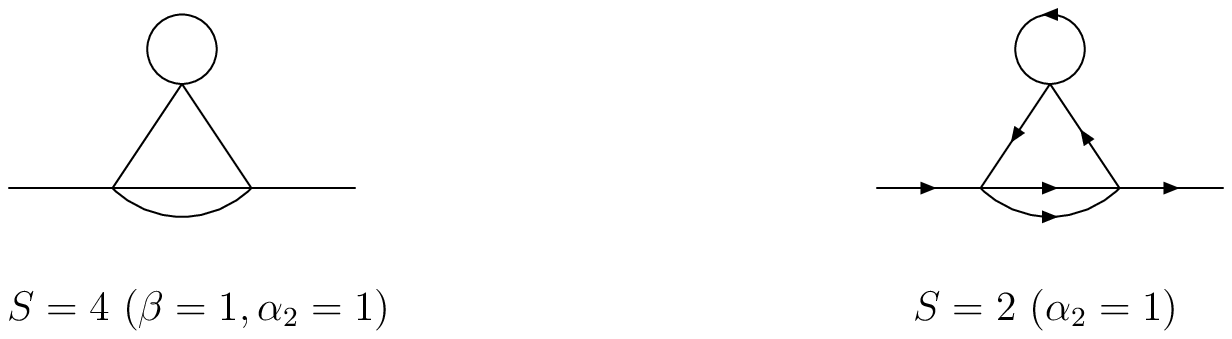}
\includegraphics{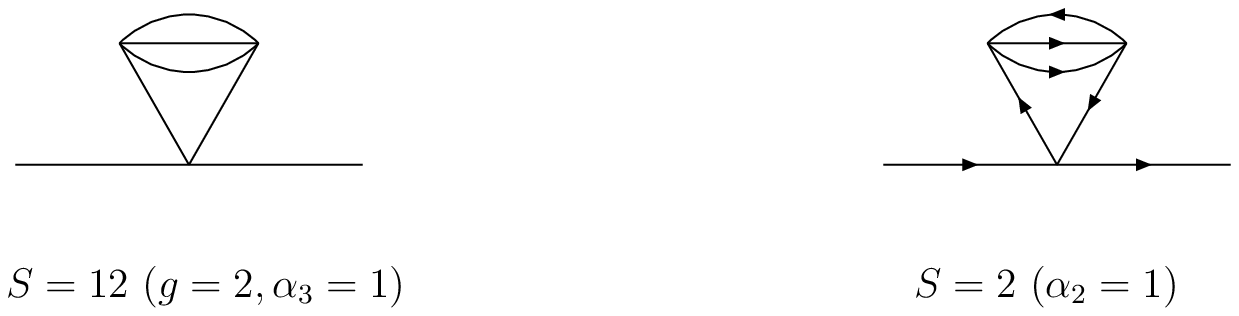}
\ec
\end{figure}

\begin{figure}[h]
\bc
\includegraphics{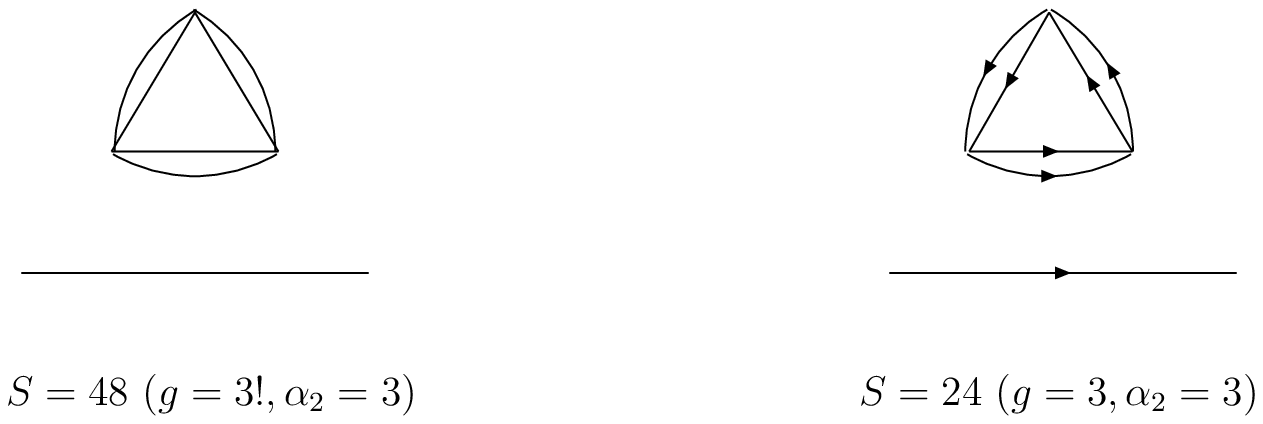}
\includegraphics{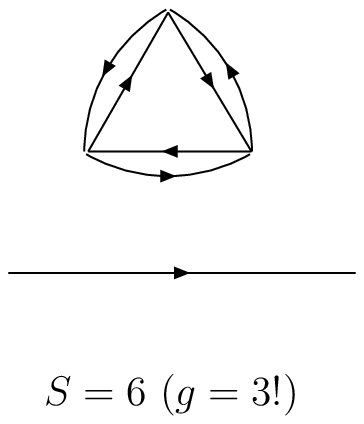}
\includegraphics{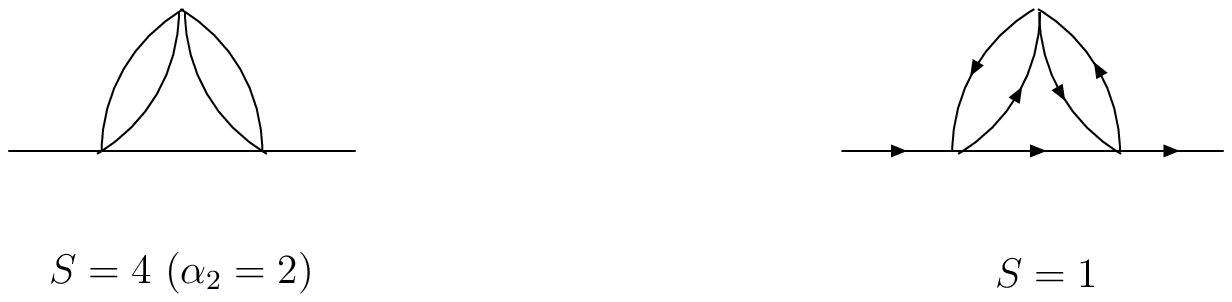}
\includegraphics{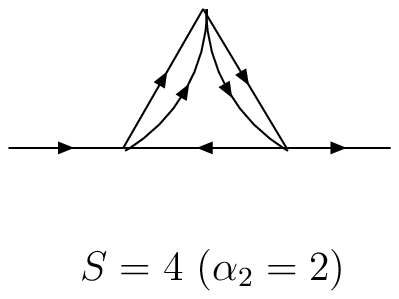}
\ec
\end{figure}

\end{document}